\documentclass[12pt]{article}
\usepackage{psfig}

\def\refb#1{(\ref{#1})}
\newcommand{\be}{\begin{equation}}
\newcommand{\ee}{\end{equation}}
\newcommand{\ba}{\begin{eqnarray}}
\newcommand{\ea}{\end{eqnarray}}

\def\laq{\raise 0.4ex\hbox{$<$}\kern -0.8em\lower 0.62ex\hbox{$\sim$}}
\def\gaq{\raise 0.4ex\hbox{$>$}\kern -0.7em\lower 0.62ex\hbox{$\sim$}}
\begin{document}
\title{A New Era in High-energy Physics}
\author{Eun-Joo Ahn\footnote{Email: sein@oddjob.uchicago.edu}\\
\small Department of Astronomy and Astrophysics, University of Chicago,\\
\small 5640 S.\ Ellis Avenue, Chicago, IL 60637 USA\\ \\
Marco Cavagli\`a\footnote{Email: cavaglia@mitlns.mit.edu}\\
\small Center for Theoretical Physics,
\small Massachusetts Institute of Technology,\\
\small77 Massachusetts Avenue, Cambridge MA
02139-4307 USA}
\date{(MIT-CTP-3255. \today)}
\maketitle
\begin{abstract}
In TeV-scale gravity, scattering of particles with center-of-mass energy
of the order of a few TeV can lead to the creation of nonperturbative,
extended, higher-dimensional gravitational objects: 
{\it Branes}. Neutral or charged, spinning or spinless,
Einsteinian or supersymmetric, low-energy branes could dramatically
change our picture of high-energy physics. Will we create branes in
future particle colliders, observe them from ultra high energy cosmic rays,
and discover them to be dark matter?
\\\\
{\it Essay written for the awards of the 2002 Gravity Research Foundation
competition.}
\end{abstract}

\newpage 
We may be on the verge of a new and unexpected era in high-energy
physics. Two different fundamental energy scales are observed in nature:
the electroweak scale, $E_{EW}\sim 1$ TeV, and the gravitational scale,
$E_{G}\sim 10^{16}$ TeV. Unification of these two scales must be encoded in any
Grand Unification Theory: either the gravitational scale is lowered to the
electroweak scale by some unknown physics \cite{Antoniadis:1990ew}, or vice
versa. In the latter, nonperturbative ``quantum'' gravity effects become
apparent at energy scales of the order $E_G$, whereas in the former
gravitational phenomena become strong at energies sixteen orders of magnitude
lower. If the fundamental scale is of the order of TeV, a collision of two
particles with center-of-mass energy larger than a few TeV may lead to the
formation of gravitational objects, such as black holes \cite{Banks:1999gd}.
Nonperturbative gravitational phenomena would be observed in any physical
process with energy above the TeV scale. The early universe, cosmic rays
\cite{Feng:2001ib} and particle colliders \cite{Giddings:2001bu} are a few
examples. 

String theory \cite{Polchinski:book} has emerged as the most successful
candidate for the theory of quantum gravity. The five consistent superstring
theories and eleven-dimensional supergravity are connected by a web of duality
transformations and constitute special points of a multi-dimensional moduli
space of a more fundamental, nonperturbative (M-)theory. In addition to
strings, nonperturbative formulation of string theory contains
higher-dimensional, nonperturbative, extended objects called {\it branes}
\cite{Stelle:nv}. TeV-scale gravity can be naturally realized in string theory
(see, e.g., \cite{Antoniadis:2001sw}). Therefore, if string theory is the
ultimate theory of nature, and the Planck scale is of the order TeV, a plethora
of non-perturbative processes are possible at the TeV scale: In addition to
black holes, branes will form as well. It should be noted that brane formation
is a generic phenomenon that happens in any  gravitational theory. The presence
of a  number of extra-dimensions is sufficient to allow for the existence of
extended objects, though the phenomenology of the creation and decay of branes 
may depend on the theory. In this essay we propose that branes are created by
super-Planckian scattering processes in TeV-scale gravity, and discuss some
phenomenological implications of brane formation in string and Einstein
theories.

A $p$-dimensional non-spinning extended object propagating in a $D$-dimensional
spacetime is described by the metric 
\be
ds^2=R(r)^{a_1}(-dt^2+\delta_{ij}dy^idy^j)+R(r)^{a_2}dr^2+r^2R(r)^{a_3}d\Omega^2_{D-p-2}\,,
\label{pbrane}
\ee
where $y_i$ ($i,j=1,\dots,p$) are the brane coordinates, $d\Omega^2$ is the
line element of the $(D-p-2)$-dimensional unit sphere, and
\be
R(r)=1-\left({r_{p}\over r}\right)^{D-p-3}\,.
\label{R}
\ee
The explicit value of the parameters $a_i$ depends on the underlying
gravitational theory. Here, we focus on $D$-dimensional Einstein gravity and,
motivated by M-Theory, eleven-dimensional supergravity. In the former, the
parameters $a_i$ are \cite{Ahn:2002mj}
\be
a_1={\Delta\over p+1}\,,\qquad a_2={2-q-\Delta\over
q-1}\,,\qquad a_3={1-\Delta\over q-1}\,,
\ee
where $q=D-p-2$ and
\be
\Delta=\sqrt{{q(p+1)\over p+q}}\,.
\ee
Eleven-dimensional supergravity admits an elementary/electric two-brane and a
mag\-ne\-tic/so\-litonic five-brane \cite{Stelle:nv}. The parameters $a_i$ are
\be
Two-brane:\qquad a_1=2/3\,,\qquad a_2=-2\,,\qquad a_3=0\,,
\ee
\be
Five-brane:\qquad a_1=1/3\,,\qquad a_2=-2\,,\qquad a_3=0\,.
\ee
In analogy to the black hole case \cite{Banks:1999gd}, scattering of two
partons with impact parameter $b\,\laq\, r_{p}$ produces a $p$-brane
described by a suitable localized energy field configuration and whose exterior
is described by Eq.\ \refb{pbrane}. The cross section for the process depends
on the brane tension and is given by the geometrical cross section
corresponding to the black absorptive disk of radius $r_p$ \cite{Ahn:2002mj}. In
fundamental units, the cross section is given by  
\be
\sigma_i\sim\pi r_{p,i}^2=F_i(n,p)V_p^{-{2\over n-p+1}}s^{1\over n-p+1}\,.
\label{sigma}
\ee
where $s=E_{ij}^2$ is the square of the center-of-mass energy of the two
scattering partons, and $V_p$ is the volume of the brane.  The form factor
$F_{i}(n,p)$ depends on the model considered. For the Einsteinian brane we find
\be
\displaystyle
F_E(n,p)=\left[64(p+1)\,\Gamma \left[(n+3-p)/2\right]^2
\over (2+n)(n-p+2)\right]^{1\over n-p+1} \,.
\label{ff-Einstein}
\ee
The electric and magnetic supergravity branes have
\be
F_{el}(n,p)=2\,,\qquad F_{mg}(n,p)=(2\sqrt{\pi})^{2/3}\,,
\ee
respectively.  The total cross-section for a generic scattering process can be
calculated from Eq.\ \refb{sigma}. The Large Hadron Collider (LHC) with a
proton-proton center-of-mass energy of $14$ TeV will possibly offer the first
opportunity to observe brane formation. Assuming a fundamental Planck scale of
$M_{\star}=2$ TeV, and $D=10$ dimensions, the total cross sections for the
formation of Einsteinian branes at LHC is plotted in
Fig.~1.

\null\hskip 0.1truein\psfig{file=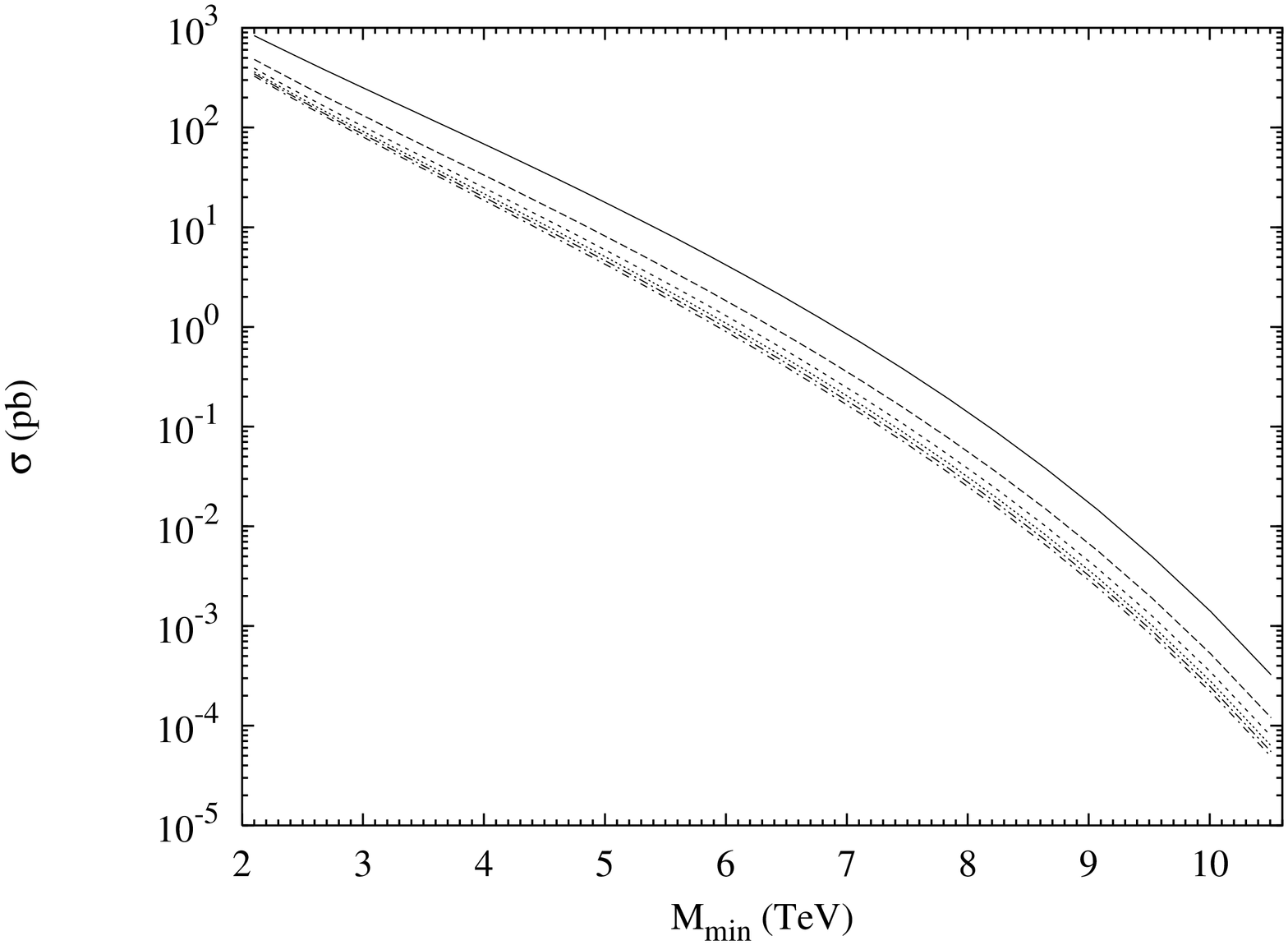,angle=270,width=5truein}

{\small\noindent Fig.\ 1: Cross section (pb) for the formation of branes more
massive than M$_{\rm min}$ (TeV) at LHC ($p=0\dots 5$ from below). The volume
of the branes is assumed to be equal to one in fundamental Planck units.}

For this particular choice of parameters the cross sections for brane
production at LHC are in the range $10^{-4}-10^{3}$ pb. The cross section
increases for increasing brane dimension. Therefore, formation of
higher-dimensional branes dominates formation of lower-dimensional branes and
spherically symmetric black holes ($0$-branes). For a minimum brane mass of
$M_{\rm min}=3$ TeV, the cross section for a formation of a five- and a
two-Einsteinian brane is $\sigma_5\approx 250$ pb and $\sigma_2\approx 90$ pb,
respectively. Therefore, with a LHC luminosity of $L=3\cdot 10^{4}~{\rm
pb}^{-1}~{\rm yr}^{-1}$ we expect a five-brane event and a two-brane event
approximately every 5 and 10 seconds.  

Production of branes at particle colliders - if observed - would allow to
investigate the structure of the extra-dimensions. Brane cross sections are
very sensitive to the size of the brane, which is related to the size of the
compactified extra dimensions around which the brane wraps. The cross section
is enhanced if the length of the extra dimensions is sub-Planckian. For
instance, the cross section of a five-brane wrapped on extra dimensions with
size 1/2 of the fundamental scale is enhanced by a factor $\approx 10$. An
enhancement of the cross section would have important consequences in
high-energy cosmic ray physics, since a sufficient flux of $p$-branes could be
detected by ground array and air fluorescence detectors \cite{Jain:2002kf}. 

We expect the creation of bosonic non-supersymmetric (non-BPS \cite{Stelle:nv})
branes in particle colliders and high-energy cosmic rays.  Although the decay
process of a bosonic brane is not understood, string field theory suggests that
a higher-dimensional brane can be seen as a lump of lower-dimensional branes
\cite{Sen:1999mh}. The tension of the brane causes the latter to decay in lower
dimensional branes, and eventually to evaporate as a black hole. Therefore, a
bosonic non-supersymmetric brane can be considered as an intermediate state in
the scattering process. Pursuing the analogy with particle physics, black holes
can be regarded as a metastable particles and branes their
resonances.\footnote{We are grateful to Angela Olinto for this remark.} 

Extremal supersymmetric branes of eleven-dimensional supergravity, however,
saturate the Bogomol'ny bound, have zero entropy, and do not evaporate
\cite{Duff:1996hp}. Therefore, if supersymmetry is unbroken and
eleven-dimensional supergravity describes the physics at energies above the TeV
scale, high-energy particle scattering produces stable branes (Fig.\ 2). In the
standard cosmological scenario \cite{Kolb:Book} and in the new brane-world
cosmological models \cite{Khoury:2001wf}, the temperature of the early universe
is expected to have reached super-TeV values: Creation of BPS branes could have
been a common event in the early universe. At temperatures above the
fundamental scale we expect a plasma of branes in thermal equilibrium with the
primordial bath. At temperatures of the order of TeV, branes decouple from the
thermal plasma, leaving stable BPS relics. Today these relics would appear to
an observer like heavy supersymmetric particles with mass $M_{br}\sim$ TeV and
cross sections $\sigma_{br}\laq$ pb, thus providing a candidate for dark
matter. Most interestingly, a gas of branes leads to a cosmological model that
solves the initial singularity and horizon problems of the standard
cosmological model without relying on an inflationary phase
\cite{Alexander:2000xv}.  

\null\hskip 0.1truein\psfig{file=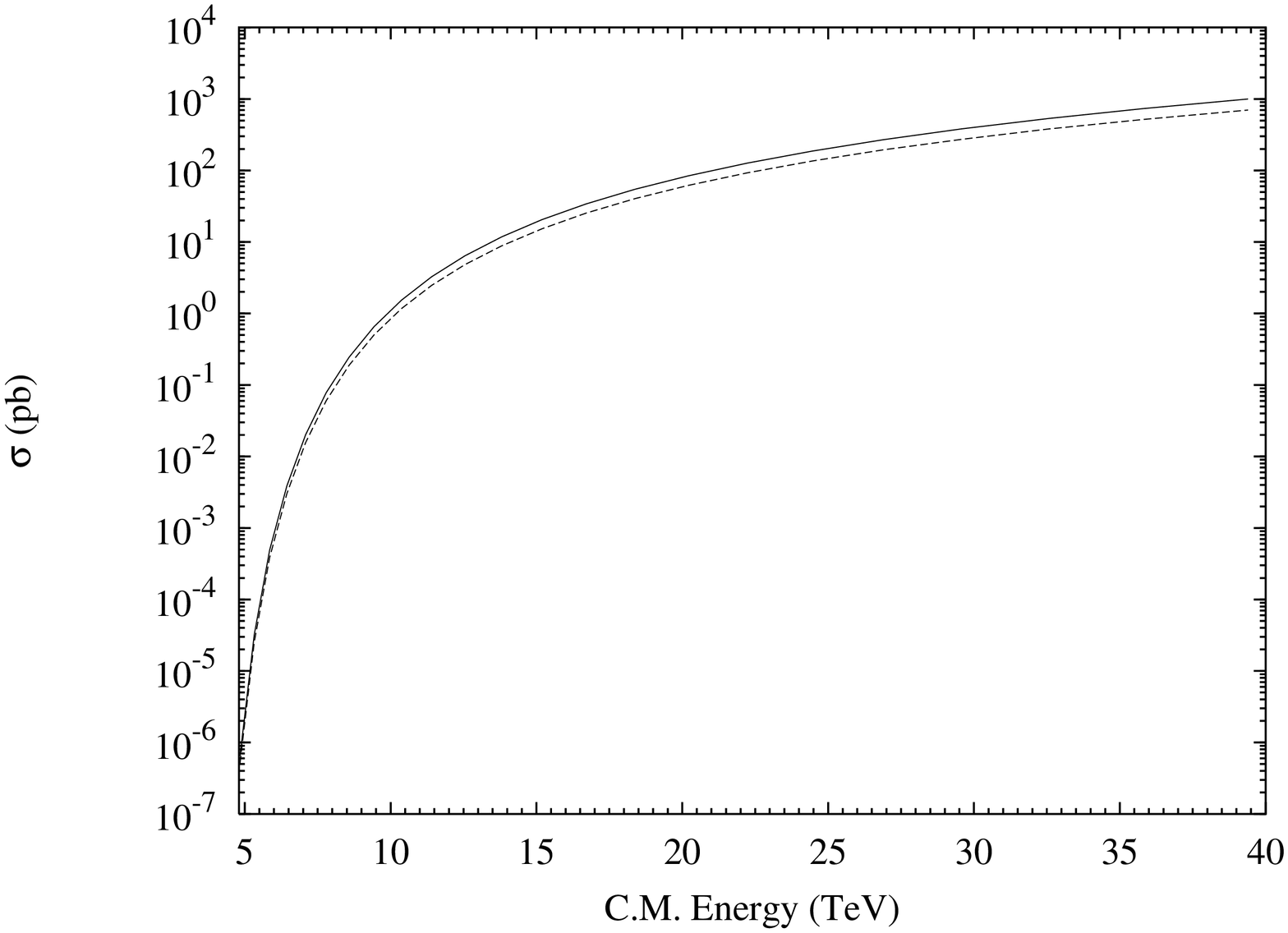,angle=270,width=5truein}

{\small\noindent Fig.\ 2: Cross sections for creation of electric (lower) and
magnetic (upper) supergravity branes by proton-proton scattering with
$M_\star=2$ TeV, minimum mass $M_{\rm min}=2M_\star$, and unit volume. The
cross sections for the corresponding Einsteinian two- and five-brane are
enhanced by a factor of $\sim 2.2$ and $\sim 2.7$, respectively.}

To conclude, in gravitational theories with large-extra dimensions creation of
nonperturbative extended objects is expected to happen and have important
effects on all physical processes above the TeV scale. Non-BPS brane formation
in particle colliders and in the atmosphere by ultra high energy cosmic rays
will probe short-distance physics and the structure of the extra dimensions. In
cosmology, primordial creation of stable BPS branes may have played an
important role in the dynamics of the very early universe. Brane relics could
be the dark matter that is observed today. If the large-extra dimension
scenario does really describes the material world and its phenomena, we may
well be on the verge of a new and unexpected era in high-energy physics. 

\vskip 1em
\leftline{\large\bf Acknowledgements}

\noindent
We are very grateful to I.~Antoniadis, U.~d'Alesio, G.~Dvali, J.~Feng,
C.~Giunti, H.~Goldberg, A.~Hanany, B.C.~Harms, G.~Karatheodoris, A.~Olinto,
J.~Polchinski, and B.~Zwiebach, for interesting discussions and useful
comments. E.-J.~A.\ and M.~C.\ thank MIT and UoC for the kind hospitality,
respectively. This work is supported in part by funds provided by the U.S.\
Department of Energy under cooperative research agreement DE-FC02-94ER40818.

\end{document}